\documentclass[aps,prd,12point,twocolumn,nofootinbib,showpacs,superscriptaddress]{revtex4-1}
\def\theequation{\arabic{section}.\arabic{equation}}
\usepackage{psfrag}
\usepackage{subfigure}
\usepackage{color}
\usepackage{mathrsfs}
\usepackage{graphicx}
\usepackage{amssymb, bm}
\usepackage{amsmath, amsthm}
\usepackage{epstopdf}
\usepackage{hyperref}
\usepackage{enumerate}
\usepackage{longtable}

\newcommand{\be}{\begin{equation}}
\newcommand{\ee}{\end{equation}}

\begin{document}
\def\theequation{\arabic{section}.\arabic{equation}} 
\title{Two new approaches to the anomalous limit of Brans-Dicke to 
Einstein gravity}
\author{Valerio Faraoni}
\email[]{vfaraoni@ubishops.ca}
%\thanks{}
\affiliation{Department of Physics and Astronomy and {\em STAR} Research 
Cluster, Bishop's University, 2600 College Street, Sherbrooke, Qu\'ebec, 
Canada J1M~1Z7 }

\author{Jeremy C\^ot\'e}
\email[]{jcote16@ubishops.ca}
%\thanks{}
\affiliation{Department of Physics and Astronomy, Bishop's University, 
2600 College Street, Sherbrooke, Qu\'ebec, 
Canada J1M~1Z7}

%\date{\today}
\begin{abstract}

Contrary to common belief, (electro)vacuum Brans-Dicke gravity does not 
reduce to general relativity for large Brans-Dicke coupling $\omega$, a 
problem which has never been fully solved. Two new approaches, independent 
from each other, shed light on this issue producing the same result: in 
the limit $\omega \rightarrow \infty$ an (electro)vacuum Brans-Dicke 
spacetime reduces to a solution of the Einstein equations sourced, not by 
(electro)vacuum, but by a minimally coupled scalar field. The latter is 
shown to coincide with the Einstein frame scalar field. The first method 
employs a direct analysis of the Einstein frame, while the second 
(complementary and independent) method uses an imperfect fluid 
representation of Brans-Dicke gravity together with a little known 
1-parameter symmetry group of this theory.

\end{abstract}
%\keywords{}

\maketitle

\section{Introduction}
\label{sec:1}
\setcounter{equation}{0}

There is currently a major theoretical and experimental effort to detect, 
or constrain, deviations of gravity from Einstein's general relativity 
(GR), which span many areas of research and many spatial scales 
\cite{tests, Padilla, Psaltis}, including 
cosmology \cite{Koyama, Euclid}, supermassive black holes 
\cite{smBHtests}, stellar mass black 
hole binaries emitting gravitational waves \cite{gwtests}, and Solar 
System tests 
\cite{Reasenberg, BertottiIessTortora, Will, Everitt, Nature, tests, 
Padilla, Psaltis}. The 
prototypical 
alternative to GR is scalar-tensor gravity \cite{BransDicke, ST, 
Nordtvedt}, which is motivated by fundamental theoretical considerations 
and by observational cosmology. On the one hand, every attempt to quantize 
gravity produces deviations from GR in the form of higher order equations 
of motion, extra gravitational degrees of freedom, or curvature 
corrections to the Einstein-Hilbert action (for example, the low-energy 
limit of bosonic string theory, the simplest string theory, yields a 
Brans-Dicke gravity with coupling parameter $\omega=-1$ \cite{bosonic}). 
On 
the other hand, from the cosmological point of view, explaining the 
present acceleration of the universe without invoking an {\em ad hoc} dark 
energy \cite{AmendolaTsujikawabook} has led to the very popular $f(R)$ 
class of theories \cite{CCT} (where $R$ is the Ricci scalar of spacetime). 
This is nothing but a subclass of Brans-Dicke theories in disguise, with 
coupling parameter $\omega=0$ and a complicated scalar field potential 
(see Refs. 
\cite{reviews} for reviews). Also the most successful model of inflation 
in the early universe, Starobinsky inflation, is based on quadratic 
curvature corrections to Einstein gravity, embodied in the $f(R)=R+\alpha 
R^2$ Lagrangian \cite{Starobinsky}.

It is assumed that viable modified theories of gravity have some limit to 
GR and 
observational tests then tell us that gravity must indeed be close to GR 
on the scales at which such tests are available (which, admittingly, do 
not span a vast range \cite{tests}). Possessing a GR limit seems to be an 
essential ingredient 
of the required ``closeness to GR'', but there are still gaps in our 
theoretical understanding of this limit even for the simplest alternative 
to GR and the simplest incarnation of scalar-tensor gravity, which is the 
original Brans-Dicke theory \cite{BransDicke}. Here we clarify a puzzle 
in the limit of vacuum Brans-Dicke gravity (which is relevant for Solar 
System tests) to GR.

The scalar-tensor action is (we follow the 
notation of Ref.~\cite{Waldbook}, using units in which Newton's 
constant $G$ and the speed of light $c$ are unity) 
\begin{eqnarray}
S_{ST} &=& \frac{1}{16\pi} \int d^4x \sqrt{-g} \left[ \phi R 
-\frac{\omega(\phi )}{\phi} 
\, \nabla^c\phi \nabla_c\phi -V(\phi) \right] \nonumber\\
&&\nonumber\\
&\, & +S^{(m)} \,, \label{STaction}
\end{eqnarray}
where $\phi>0$ is the Brans-Dicke scalar field,  the function 
$\omega(\phi)$ 
(a constant parameter in the original Brans-Dicke theory 
\cite{BransDicke}) is the ``Brans-Dicke coupling'', $V(\phi)$ is a 
scalar field potential, and $S^{(m)}=\int d^4x \sqrt{-g} \, {\cal L}^{(m)} 
$ is the matter part of the action.

The variation of the action~(\ref{STaction}) with respect to the inverse 
metric $g^{ab}$ and to 
the Brans-Dicke scalar $\phi$ yields the (Jordan frame) field equations  
\cite{BransDicke, ST} 
\begin{eqnarray}
R_{ab} - \frac{1}{2}\, g_{ab} R &=& \frac{8\pi}{\phi} \,  T_{ab}^{(m)} 
\nonumber\\
&&\nonumber\\
&\, & + \frac{\omega}{\phi^2} \left( \nabla_a \phi 
\nabla_b \phi -\frac{1}{2} \, g_{ab} 
\nabla_c \phi \nabla^c \phi \right) \nonumber\\
&&\nonumber\\
&\, &  +\frac{1}{\phi} \left( \nabla_a \nabla_b \phi 
- g_{ab} \Box \phi \right) 
-\frac{V}{2\phi}\, 
g_{ab} \,,\nonumber\\
&& \label{BDfe1}\\
\Box \phi = \frac{1}{2\omega+3} & & 
\left( \frac{8\pi T^{(m)} }{\phi}   + \phi \, \frac{d V}{d\phi} 
-2V -\frac{d\omega}{d\phi} \nabla^c \phi \nabla_c \phi \right) \nonumber\\ 
&& \label{BDfe2}
\end{eqnarray}
where $R_{ab}$ is the Ricci scalar and $\nabla_a $ is the covariant 
derivative of the spacetime metric $g_{ab}$,  while $ T^{(m)} \equiv 
g^{cd}T_{cd}^{(m)} $ is the trace of 
the matter energy-momentum tensor $T_{ab}^{(m)}=-\frac{2}{\sqrt{-g}} \, 
\frac{\delta S^{(m)}}{\delta g^{ab}} $. The matter stress-energy tensor 
and the effective stress-energy tensor of the scalar $\phi$ are 
covariantly conserved 
separately.

Let us briefly review the limit to GR of Brans-Dicke theory and its 
anomaly. GR is reproduced from Jordan frame Brans-Dicke gravity when 
$\phi$ becomes constant: then  
the effective gravitational coupling $G_{eff}\simeq \phi^{-1}$ also 
becomes constant (for more general scalar-tensor theories in 
which $\omega=\omega(\phi)$, the limit to GR is  $\omega 
\rightarrow \infty$ in conjunction with $ \omega^{-3} d\omega/d\phi  
\rightarrow 0 $ \cite{Nordtvedt, Homann}). The contentious issue is how 
fast 
$\phi$ approaches a 
constant. It is commonly believed ({\em e.g.}, \cite{Weinberg}) 
that (Jordan frame) Brans-Dicke gravity 
reduces to GR as the Brans-Dicke parameter $\omega \rightarrow 
\infty$, with the Brans-Dicke scalar $\phi$ following the 
asymptotics 
\be  \label{1}
\phi=\phi_{\infty}+ \mathcal{O}\left( \frac{1}{\omega} \right)  
\,,
\ee
where $\phi_{\infty} >0$ is a constant. However, a number of 
analytic solutions of the Brans-Dicke field equations 
have been reported which fail to reduce to the corresponding solutions of 
GR as $\omega \rightarrow \infty$ \cite{failure, BanerjeeSen97} (including 
anomalies in matter different from (electro)vacuum 
\cite{Chauvineau, newpapers, Brandoetal2018}, 
to which we will instead restrict). In these situations, the 
asymptotic behavior of the Brans-Dicke scalar is not the one expected 
(Eq.~(\ref{1})) but\footnote{See also the discussion of  
Ref.~\cite{BhadraNandi}.} 
\be  \label{2} 
\phi=\phi_{\infty}+ \mathcal{O}\left( \frac{1}{\sqrt{\omega}} 
\right) \,.
\ee
It has also been realized that the asymptotic 
behavior~(\ref{2}) of the Brans-Dicke field is 
usually accompanied by a vanishing trace $T^{(m)} $ of the 
matter 
energy-momentum tensor  
\cite{BanerjeeSen97} (this condition is trivially satisfied {\it in 
vacuo}). This coincidence hints to some 
degree of conformal invariance, which has motivated an explanation of 
why the $\omega \rightarrow \infty$ 
limit of Brans-Dicke theory fails to reproduce GR {\em in 
vacuo} or in electrovacuo \cite{myBDlimit, mycosmobook}. We summarize 
this explanation in Sec.~\ref{sec:4}. 

The limit to GR is important for three reasons. First, it 
is related to the 
weak-field limit of gravity, in which the relativistic corrections to 
Newtonian gravity are parametrized by the so-called PPN formalism 
\cite{Will}. This formalism is the 
basis for constraining $\omega$ with Solar 
System experiments \cite{BertottiIessTortora, Will}. Second, various 
authors have 
studied attractor mechanisms in 
which scalar-tensor gravity converges toward GR, such as during the 
early evolution of the universe \cite{convergence, convergence2}. Third, 
scalar-tensor gravity could be an ``excitation'' of GR in the context of 
the thermodynamics of spacetime, in which the Einstein equations are 
derived as a sort of macroscopic equation of state \cite{Jacobson1}. Then,  
GR would represent a ``state of equilibrium'' while deviations from it 
(for  example, through the excitation of other gravitational scalar 
degrees of freedom such as the Brans-Dicke scalar $\phi$) could be 
non-equilibrium states \cite{Jacobson2, ChircoLiberati}.

Here we revisit the anomaly in the $\omega\rightarrow \infty $ limit 
of Brans-Dicke theory with two new approaches. The first one consists of 
using the Einstein conformal frame, while the second describes  
the Brans-Dicke field equations as effective Einstein equations with
an effective imperfect fluid made of terms including derivatives of the 
scalar field $\phi$. Both approaches produce the same result.

%%%%%%%%%%%%

\section{Einstein frame approach}
\label{sec:2}
\setcounter{equation}{0}

The first approach relies on the fact that scalar-tensor gravity has 
another close relation with GR, besides the $\omega\rightarrow \infty$ 
limit. Let us restrict, for simplicity, to {\em vacuum} Brans-Dicke 
theory: the Einstein frame formulation of this theory is formally GR with 
a metric 
$\tilde{g}_{ab}$ and a (minimally coupled and canonical) scalar field 
$\tilde{\phi}$. So the question arises naturally: what is the relation 
between the $\omega\rightarrow \infty$ limit $\left( g_{ab}^{(\infty)}, 
\phi^{(\infty)} \right)$ of a Brans-Dicke spacetime $\left( g_{ab}, \phi 
\right)$ and its Einstein frame version $\left( \tilde{g}_{ab}, 
\tilde{\phi}\right)$? One could expect these two to coincide, 
but they do not, as is elucidated in the following.

\subsection{Einstein frame}

In addition to the Jordan frame $\left( g_{ab}, \phi 
\right)$, another representation of scalar-tensor 
gravity, the Einstein frame $\left( 
\tilde{g}_{ab}, \tilde{\phi} \right)$, is used \cite{Dicke}. 
The metric tensors in the Einstein and in the Jordan frames 
are related by the conformal transformation 
\be \label{metrictransformation}
g_{ab} \rightarrow \tilde{g}_{ab} \equiv \phi \, g_{ab} \,,
\ee
while, for the two scalar fields, we have 
\be
d\tilde{\phi} = \sqrt{ \frac{|2\omega+3|}{16\pi}} \, \frac{d\phi}{\phi} 
\,.
\ee
Restricting ourselves to Brans-Dicke theory with 
constant $\omega$, the scalar field is redefined 
non-linearly as 
\be
\phi \rightarrow \tilde{\phi}=\sqrt{\frac{|2\omega+3|}{16
\pi}} \, \ln \left( \frac{\phi}{\phi_0}\right) \,, 
\label{phitransform}
\ee
where $\phi_0$ is an integration constant and $\omega\neq -3/2$. Since 
both 
$\phi$ and $g_{ab}$ depend on the parameter $\omega$, barring miraculous 
cancellations, in general the 
Einstein frame metric $\tilde{g}_{ab}$ given by 
Eq.~(\ref{metrictransformation}) depends on $\omega$. Similarly, the  
Einstein frame scalar $\tilde{\phi}$ given by Eq.~(\ref{phitransform})  
depends on $\omega$.

In the Einstein frame, the Brans-Dicke action ({\em i.e.}, 
(\ref{STaction}) with $\omega=$~const.) 
assumes the form
\be
S_\text{BD} = \int d^4x\sqrt{-\tilde{g}}\left[
\frac{\tilde{R }}{16 \pi}-\frac{1}{2} \, 
\tilde{g}^{ab}\nabla_a\tilde{\phi} \nabla_b\tilde{\phi}  
-U(\tilde{\phi})  + \frac{{\cal  
L}^{(m)}}{\phi^2(\tilde{\phi}) }  \right]  \,, 
\label{BDactionEframe} 
\ee
where
\be
U (\tilde{\phi}) = 
\frac{V(\phi)}{16\pi \phi^2}\left|_{\phi=\phi( \tilde{\phi}) } 
\right.  \label{ppotential} 
\ee
(Einstein frame quantities are denoted by a 
tilde). The action~(\ref{BDactionEframe}) is formally the 
Einstein-Hilbert action of GR with a matter scalar field 
which has canonical kinetic energy density, except that 
now this scalar couples non-minimally to matter. The 
Einstein frame field equations are
\begin{eqnarray}
\tilde{R}_{ab}-\frac{1}{2} \, \tilde{g}_{ab} \tilde{R}  &=& 
8\pi \left(  \mbox{e}^{- \sqrt{\frac{64\pi}{|2\omega+3|}} 
\, \tilde{\phi} } \, T_{ab}^{(m)} + 
\tilde{\nabla}_a \tilde{\phi} \tilde{\nabla}_b
\tilde{\phi} \right.\nonumber\\
&&\nonumber\\
&\, & \left.
-\frac{1}{2} \, \tilde{g}_{ab}\, 
\tilde{g}^{cd} \tilde{\nabla}_c \tilde{\phi} \tilde{\nabla}_d \tilde{\phi} 
 - U(\tilde{\phi})  \, \tilde{g}_{ab} \right)   \,,\nonumber\\ 
&& \label{Eframefe}
\end{eqnarray}
\be
 \tilde{g}^{ab} \tilde{ \nabla}_a  \tilde{ \nabla}_b
\tilde{\phi} -\frac{dU}{d\tilde{\phi}} 
+8\, \sqrt{ \frac{\pi}{|2\omega+3|}} \,  \mbox{e}^{ 
-\sqrt{\frac{64\pi}{|2\omega+3|}}\, \tilde{\phi} } \, {\cal 
L}^{(m)} = 0 \,. 
\label{EframeKG}
\ee
Let us restrict, for simplicity, to vacuum Brans-Dicke 
theory by setting $T_{ab}^{(m)}=0$.  Then the 
explicit coupling 
between the Einstein frame scalar $\tilde{\phi}$ and 
matter disappears from the action and from the field equations  
and the Einstein frame 
action~(\ref{BDactionEframe})  
formally reduces to the Einstein-Hilbert action of GR. Therefore, the 
Einstein frame pair $\left( \tilde{g}_{ab}, \tilde{\phi} \right)$ 
is formally a scalar field  solution of GR corresponding to the original 
Jordan 
frame spacetime  $\left( g_{ab}, \phi \right)$ and it could appear as a 
natural candidate for a ``GR limit'' of the latter.

\subsection{Comparison between the Einstein frame and the 
$\omega\rightarrow \infty$ limit of the Jordan frame}

Since we are interested in the anomaly discussed in the literature 
for the limit $\omega\rightarrow \infty$ of Brans-Dicke theory, we are 
only interested in the vacuum case in the following.

As already seen, in the Einstein frame both the metric $\tilde{g}_{ab}$ 
and the scalar 
$\tilde{\phi}$ depend on the parameter $\omega$, but the gravitational 
coupling is constant. One cannot regard the Einstein frame fields $ \left( 
\tilde{g}_{ab}, 
\tilde{\phi} \right)$ as  a limit to GR of $\left( g_{ab}, \phi \right)$ 
because 
any dependence on the parameter $\omega$ (which is absent in GR) must 
disappear after taking the GR limit. Moreover,  $\tilde{g}_{ab}$ is a 
solution of 
the coupled Einstein-Klein-Gordon equations with Klein-Gordon field 
$\tilde{\phi}$. There are different matter sources in the two conformal 
frames: vacuum in the Jordan frame and a Klein-Gordon field in the 
Einstein frame, and the GR limit of a Brans-Dicke solution must have the 
same matter source.\footnote{$\phi$ has a gravitational nature in the 
Jordan frame but it appears as a matter field in the Einstein frame, 
blurring the sharp distinction between geometry and matter present in 
Einstein theory. Ambiguities in the identification of matter
and gravitational fields are an obstacle to creating a meta-theory of 
gravitational theories \cite{SotiriouLiberatiFaraoni2008}.}  It is more 
meaningful to compare 
Einstein and Jordan frame in the $\omega \rightarrow \infty$ limit, as 
is done in the next subsection.

\subsection{Comparison between the $\omega\rightarrow \infty$ limits   
of the Einstein and Jordan frames}

The $\omega\rightarrow \infty$ limit of a Jordan frame 
spacetime which is a solution of the Brans-Dicke field 
equations~(\ref{BDfe1}) and~(\ref{BDfe2}) 
is expected to produce\footnote{The constant $\phi_{\infty}$ 
is positive because it corresponds to the inverse of the gravitational 
coupling strength.} $\phi_{\infty} = 
$~const.~$>0$ and the limit $g_{ab}^{(\infty)}$ of the metric, 
while the scalar field 
potential reduces to an effective  cosmological constant 
$\Lambda \equiv V(\phi_{\infty})/(2\phi_{\infty})$. The conformal 
map relating the Jordan and Einstein frame metrics yields, in this 
limit, 
\be
\tilde{g}_{ab}=\phi \, g_{ab} \;
 \longrightarrow  \; \phi_{\infty} \, g_{ab}^{(\infty)} 
\;\;\;\;\mbox{as} \;\; \omega \rightarrow
\infty \,.
\ee
By dropping the irrelevant multiplicative 
constant $\phi_{\infty}$, which can always be absorbed by a 
coordinate redefinition, one has   
$\tilde{g}_{ab}(\omega) 
\rightarrow g_{ab}^{(\infty)}$ in this limit. 
Therefore, the Einstein 
frame {\em geometry} always coincides with the $\omega 
\rightarrow \infty$ limit of the Jordan frame {\em 
geometry}. The Einstein frame scalar field is given by 
Eq.~(\ref{phitransform}).  In the limit $\omega\rightarrow \infty$, the 
square  root in the right hand side of Eq.~(\ref{phitransform}) diverges, 
then the 
entire right hand side diverges, unless $\phi$ becomes a constant 
$\phi_{\infty}$ as 
$\omega\rightarrow \infty$, which is indeed what happens. This is not 
sufficient, however, to avoid the divergence of $\tilde{\phi}$: it must 
also be 
$\phi_0=\phi_{\infty}$, which 
makes $\ln (\phi/\phi_{\infty})$ vanish so that 
$\tilde{\phi}$ has a chance to remain finite in this limit. The 
divergence of $\tilde{\phi}$ as $\omega\rightarrow \infty$ would be 
unphysical since this scalar field must be well defined for all values of 
$\omega$, including large ones (except, possibly, at spacetime 
singularities). Hence, we require $\phi_0= \phi_{\infty}$ 
and we write 
\be
\tilde{\phi}= \sqrt{ \frac{ |2\omega+3|}{16\pi} } \, \ln 
\left( \frac{\phi}{\phi_{\infty} } \right) \,. \label{100}
\ee
This argument fixes the arbitrary integration constant $\phi_0$ appearing 
in Eq.~(\ref{phitransform}).

If the Jordan frame scalar has asymptotics
\be
\phi=\phi_{\infty} +
\mathcal{O} \left( \frac{1}{\omega^{p} } \right) 
\ee
for some\footnote{Here $\mathcal{O}(\omega^{-p}) $ denotes a scalar 
function 
of the coordinates which is of order $\omega^{-p}$  as $\omega 
\rightarrow \infty$.} $p$, then the Einstein frame scalar~(\ref{100}) 
can be written as
\be
\tilde{\phi} = \frac{1}{2\sqrt{2\pi} } \, 
\frac{ \ln \left[ 1+ \mathcal{O} \left( 1/\omega^p \right)\right] 
}{
\frac{1}{\sqrt{|\omega+3/2|} } }  
\approx \frac{1}{2\sqrt{2\pi} } \,
 \mathcal{O} \left( \frac{1}{\omega^{p -1/2}} \right) 
\label{eq:beforeinesc}
\ee
using the linear expansion $\ln \left(1+x \right) \simeq x $ for 
$|x|\ll 1$. We do not want to prescribe the asymptotics of $\phi$,  and 
attempts to do so would require independent arguments but,  {\em a 
priori}, there are only three possibilities:

\begin{itemize}

\item If $ p >1/2$, in particular in the ``standard'' 
situation 
$\phi =\phi_{\infty} +  \mathcal{O} \left(1/\omega \right) $ 
corresponding to $p=1$ \cite{Jordan, Weinberg}, 
then $\tilde{\phi} \rightarrow 0$ as 
$\omega\rightarrow \infty$ and, {\em in vacuo}, no scalar field source is 
left 
in the limit of the Einstein frame field equations (except, possibly, for a 
cosmological constant if $V(\phi_{\infty})\neq 0$). 

\item If $p=1/2$, which is the ``anomalous'' situation studied 
in the literature \cite{failure, BanerjeeSen97, myBDlimit, mycosmobook},  
$\tilde{\phi}$ does not vanish but reduces to a function 
$\tilde{\phi}_{\infty}$ which does not depend on $\omega$.
Then the limit of the Einstein frame equations contains a 
(canonical) scalar field 
source  $\tilde{\phi}_{\infty}$, unless this function is constant. If 
$\tilde{\phi}_{\infty}$ is a constant instead of depending on the 
coordinates, this scalar field disappears from the field equations (except  
if there is a potential $V(\phi)$ with $V(\phi_{\infty}) \neq 0$, 
in which 
case a cosmological constant remains). It would be tempting to assume that 
$\tilde{\phi}_{\infty}$ is constant, but this assumption is unwarranted 
at this stage.

\item If $ p <1/2$ then $\tilde{\phi} \rightarrow 
\infty$ as $\omega\rightarrow +\infty$, which is 
unphysical. 
Therefore, this case is excluded and the asymptotics described by 
$ p = 1/2$ are 
truly a borderline, extreme situation. This fact is not evident from 
the standard Jordan frame analysis: one needs to analyze 
the Einstein frame to reach this conclusion.  

\end{itemize}

The analysis of the Einstein frame, and of Eq.~(\ref{100}) that 
accompanies it, selects the scaling $\sim \omega^{-1/2}$ as a critical 
behaviour which is a boundary of the possible behaviours of $\phi$ {\em in 
the Jordan frame}. This conclusion about the Jordan frame was not 
expected to come from an analysis of the Einstein frame.

Let us compare now the previous considerations with 
the asymptotics of the Jordan frame Brans-Dicke scalar. By 
inverting Eq.~(\ref{phitransform}), one 
writes the Jordan frame scalar as 
\be
\phi=\phi_0 \, \exp \left( \sqrt{ 
\frac{16\pi}{|2\omega+3|}} 
\, \tilde{\phi}  \right) \,. \label{inescapable}
\ee
The Jordan frame scalar $\phi$ depends on 
$\omega$, a property familiar from the study of exact Jordan frame 
solutions of Brans-Dicke theory, especially in cosmology 
\cite{FujiiMaedabook, mycosmobook} and in spherical symmetry  
\cite{Just, Bronnikov, Capobook, ValerioFaycalThomasAdriana}. In order 
for the $\omega\rightarrow \infty$ 
limit to not cause unphysical divergences in the Einstein frame,  
$\tilde{\phi}$ must go over to a  
scalar function  $\tilde{\phi}_{\infty}(x)$ independent of $\omega$ in 
this limit, say
\be
\tilde{\phi} (x) =\tilde{\phi}_{\infty} (x) + \mathcal{O}\left( 
\frac{1}{\omega^q } \right) \label{eq:19Ithink}
\ee
with $ q > 0 $. There are now two possibilities:

\subsubsection{ $\tilde{\phi}_{\infty} (x) \neq 0$}

If $\tilde{\phi}_{\infty} (x) \neq 0$, Eq.~(\ref{inescapable}) 
implies that 
\be
\phi(x) = \phi_0  + \mathcal{O}\left( \frac{1}{\sqrt{|\omega|}} 
\right)  \,: \label{phiasymptotics}
\ee
these are exactly the asymptotics~(\ref{2}) which are known to make the 
$\omega\rightarrow\infty$ limit fail to 
reproduce GR in the Jordan frame. Hence, a finite and nonzero Einstein 
frame  $\tilde{\phi}$ in the  $\omega \rightarrow \infty$ limit  ({\em 
i.e.}, a finite $\tilde{\phi}_{\infty}(x) \neq 0$) {\em always} implies an 
``anomalous'' scaling of the Jordan frame $\phi$  of (electro)vacuum 
Brans-Dicke theory if $\tilde{\phi}_{\infty}  \neq 0$. {\em Vice-versa}, 
using Eq.~(\ref{100}), one concludes that the ``anomalous'' Jordan frame  
scaling~(\ref{phiasymptotics}) always determines a finite non-zero 
$\tilde{\phi}_{\infty}$ in the Einstein frame, and there is a one-to-one 
correspondence between these two phenomena in the two frames.

\subsubsection{ $\tilde{\phi}_{\infty} (x) = 0$}

If instead $\tilde{\phi}_{\infty} (x) =0$, say 
$\tilde{\phi} (x) = \mathcal{O}( 1/ \omega^q ) $ with $ q>0$, then 
the comparison of Eqs.~(\ref{eq:beforeinesc}) and~(\ref{eq:19Ithink}) 
gives $ p=q+1/2 >1/2$ and it follows from Eq.~(\ref{inescapable}) 
that 
\be
\phi = \phi_0 + \mathcal{O}\left( \frac{1}{ 
\omega^{q+1/2} } \right)  \; \;\;\;\;\;\;\;\;\; (q>0).
\ee
The ``standard'' behaviour \cite{Weinberg} of 
Eq.~(\ref{1}) is reproduced by the special value  $q=1/2$. 

\subsection{General conclusions}

It is not clear from this analysis whether values of $p$ other than 
$1/2$ and $1$ are possible, but nothing seems to forbid them as long 
as one 
keeps $p>1/2$. However, such values have not appeared in the 
literature thus far. All that one can say in the Einstein frame 
approach described here is that $p$ must be larger than, or 
equal to, $1/2$ and that $ p > 1/2$ corresponds to the 
vanishing of the function $\tilde{\phi}_{\infty} $ (which does not depend 
on  $\omega$) and of the Einstein frame scalar $\tilde{\phi} $ in the 
$\omega\rightarrow \infty $ limit.

The way $\phi$ approaches a constant is detailed by the decay of its  
gradient as  $|\omega |$ becomes larger and larger, 
\be
\nabla_a \phi =  \phi_0 \sqrt{\frac{16\pi}{|2\omega+3|}} 
\, \, \exp \left(  \sqrt{\frac{16\pi}{|2\omega+3|}}  \,\tilde{\phi} 
\right)  \, \tilde{\nabla}_a \tilde{\phi} \,. \label{stucz}
\ee
Let us examine now the limit of the Jordan 
frame field equations~(\ref{BDfe1}) and, in particular, of 
the term in their right hand side
\be
A_{ab} \equiv \frac{\omega}{\phi^2} \left( \nabla_a  
\phi \nabla_b \phi -\frac{1}{2} \, g_{ab} \, \nabla^c  
\phi \nabla_c \phi \right) \,.
\ee
In the Jordan frame analysis available in the literature, 
the failure of Brans-Dicke theory to reproduce the 
expected GR limit (which corresponds to $\phi=$~const. and to the right 
hand side of the vacuum field equations equal to zero) has been traced to 
the fact that, in conjunction with the asymptotics~(\ref{2}), the tensor 
$A_{ab}$ does not vanish in the $\omega \rightarrow \infty 
$ limit but remains of order unity \cite{failure, 
BanerjeeSen97}. In fact, using Eq.~(\ref{stucz}), the tensor $A_{ab}$ 
(before taking any limit) reads 
\begin{eqnarray}
A_{ab} &=& \frac{\omega}{\phi^2} \, \phi_0^2 \,  
\mbox{e}^{ 2\sqrt{\frac{16\pi}{|2\omega+3|}} 
\,\tilde{\phi} } 
\frac{16\pi}{|2\omega+3|}  \left( \nabla_a
\tilde{\phi} \nabla_b \tilde{\phi} \right. \nonumber\\
&&\nonumber\\
&\, & \left.  -\frac{1}{2} \, 
g_{ab} 
g^{cd} \, \nabla_c \tilde{\phi} \nabla_d 
\tilde{\phi} \right) \\
&&\nonumber\\  
& = &  16\pi \mbox{sign} (\omega) \left| \frac{\omega}{2\omega+3} 
\right| 
\left( \frac{\phi_0}{\phi} \right)^2 \, \mbox{e}^{ 2\sqrt{ 
\frac{16\pi}{ |2\omega+3|} } } \left( \nabla_a\tilde{\phi} \nabla_b 
\tilde{\phi}     \right.   \nonumber\\
&&\nonumber\\
& \, & \left.     -\frac{1}{2} \, \frac{\tilde{g}_{ab} }{\phi} \, 
\phi \, \tilde{g}^{cd}  \nabla_c \tilde{\phi} \nabla_d \tilde{\phi} 
\right)   
\end{eqnarray}
and, taking the $\omega\rightarrow \infty$ limit in which 
$\phi\rightarrow \phi_0$, 
\be
A_{ab} \rightarrow  A_{ab}^{(\infty)} = 
8\pi \, \mbox{sign}(\omega) \left( \tilde{\nabla}_a
\tilde{\phi} \tilde{\nabla}_b \tilde{\phi} -\frac{1}{2} \, 
\tilde{g}_{ab} \, \tilde{g}^{cd} 
\, \tilde{\nabla}_c \tilde{\phi} \tilde{\nabla}_d \tilde{\phi} \right) \,.
\label{stucz2}
\ee 
The Einstein frame metric $\tilde{g}_{ab}^{(\infty)}$  solves the Einstein 
equations which have as a matter source the 
scalar field $\tilde{\phi}$ with canonical stress-energy tensor 
$A_{ab}^{(\infty)}$, obtained as the limit of the {\em Jordan frame} 
stress-energy tensor:
\be
\left.
8\pi \tilde{T}_{ab}[\tilde{\phi}]  
\right|_{\mbox{{\small Einstein~frame}}} = 
\left.
A_{ab}^{(\infty)} \right|_{\mbox{{\small Jordan~frame~limit}}}   
\ee
The Einstein frame scalar $\tilde{\phi}$  
is minimally coupled to the curvature and has canonical kinetic energy 
density (if $\omega>0$). If we focus only on the metric tensor and the 
spacetime geometry, 
the two candidates for a GR limit of Brans-Dicke 
theory, {\em i})~the $\omega\rightarrow \infty$ limit of the Einstein 
frame metric  and {\em ii})~the 
$\omega\rightarrow \infty$ limit of the Jordan frame metric coincide 
(apart from the irrelevant positive multiplicative constant 
$\phi_{\infty}$ which can always by eliminated by rescaling the 
coordinates).

Since this GR 
limit $\tilde{g}_{ab}^{(\infty)}=g_{ab}^{(\infty)}$  obtained with these 
two 
different methods is not a solution of the vacuum 
Einstein equations (which would require $A_{ab}^{(\infty)}$ to vanish 
identically) but solves the coupled Einstein-Klein-Gordon equations, 
the limit to GR is regarded as an anomaly. The previous discussion using 
the Einstein frame deepens our understanding of the limit to GR.

{\it Summary.} The limit $\tilde{g}_{ab}^{(\infty)} $ of the 
Einstein frame metric coincides with 
the $\omega\rightarrow \infty$ limit $g_{ab}^{(\infty)}$ of 
the Jordan frame metric $g_{ab} $, but it does not solve the Einstein 
equations 
with the same matter source $ 
-\frac{V(\phi_{\infty} )}{2\phi_{\infty} }\,  
g_{ab} \equiv -\Lambda \, g_{ab}$ and $\phi=$~const. expected to be left 
over in the 
Jordan frame equations~(\ref{BDfe1}) after the $\omega\rightarrow \infty$ 
limit. The Einstein  frame limiting metric  
$\tilde{g}_{ab}^{(\infty)}$ solves instead the Einstein equations with a 
canonical minimally coupled scalar field $\tilde{\phi}$ described by the 
stress-energy tensor 
\be
\tilde{T}_{ab} =  \tilde{\nabla}_a
\tilde{\phi} \tilde{\nabla}_b \tilde{\phi} -\frac{1}{2} \,
g_{ab}^{(\infty)} \,
g^{cd}_{(\infty)} \, \tilde{\nabla}_c \tilde{\phi} \tilde{\nabla}_d 
\tilde{\phi}  
-U(\tilde{\phi} ) \, \tilde{g}_{ab} \,.
\ee

%%%%%%%%%
\section{Examples}
\label{sec:3}
\setcounter{equation}{0}

All stationary, spherically symmetric, asymptotically flat black holes of 
Brans-Dicke theory without potential reduce to those of GR, a well-known 
no-hair theorem (which is generalized to axial symmetry, to more 
general scalar-tensor theories, and to potentials with zero minimum)  
\cite{nohair}. For these black holes, the Brans-Dicke scalar $\phi$ 
becomes trivial ({\em i.e.}, constant) outside the event horizon. Apart 
from this situation, the most general spherical, static, asymptotically 
flat solution of the vacuum Brans-Dicke equations is a 3-parameter family 
depending on parameters $\alpha_0, \beta_0$, and $\gamma$. If $\gamma\neq 
0$, this general solution is\footnote{As shown in 
Ref.~\cite{ValerioFaycalThomasAdriana}, under certain conditions this 
solution can be recast in the less general Campanelli-Lousto form 
\cite{CampanelliLousto, Vanzo}.} 
\cite{Just, Bronnikov,ValerioFaycalThomasAdriana, BhadraSarkar}
\begin{eqnarray}
ds^2 &=&  -\mbox{e}^{ (\alpha_0+\beta_0)/r } dt^2 + \mbox{e}^{ ( \beta_0-\alpha_0)/r  
}
\left( \frac{ \gamma/r }{ \sinh ( \gamma/r ) } \right)^4 dr^2  \nonumber\\
&&\nonumber\\
&\, & + \mbox{e}^{ (\beta_0-\alpha_0)/r }  
 \left(
\frac{ \gamma/r }{ \sinh( \gamma/r) } \right)^2 r^2 d\Omega_{(2)}^2 
\,, \label{new1}
\end{eqnarray}
\be
\phi (r) = \phi_0 \, \mbox{e}^{-\beta_0/r}  \,,
\;\;\;\;\;\;\; \beta_0= \frac{\sigma}{\sqrt{|2\omega+3|} }  \,,
\label{new2}
\ee
where $d\Omega_{(2)}^2=d\theta^2 +\sin^2\theta \, d\varphi^2 $ is the line 
element on the unit 2-sphere, and\footnote{This relation only holds for 
$\gamma>0$ and there is no loss of generality in choosing positive 
$\gamma$ when $\gamma\neq 0$ \cite{Wyman}.} 
\be
4\gamma^2 = \alpha_0^2+2\sigma^2 \,,\label{gas}
\ee
while $\sigma $ is a scalar charge. If $\gamma=0$, the 
general solution is instead \cite{ValerioFaycalThomasAdriana} the 
Brans class~IV spacetime \cite{Brans}
\begin{eqnarray}
ds^2 &=& -\mbox{e}^{ -2B/r }dt^2 + \mbox{e}^{ 2B(C+1)/r }
\left( dr^2+r^2 d\Omega_{(2)}^2 \right)  \,,\nonumber\\
&&\label{classIV1}\\
\phi (r) &=& \phi_0 \, \mbox{e}^{-BC/r} \,, \label{classIV2}
\end{eqnarray}
where 
\be
B=- \frac{\left(\alpha_0+\beta_0\right)}{2} \,, \;\;\;\;\;\;\;\;  C=- 
\frac{2\beta_0}{\left( \alpha_0+\beta_0 \right)}  \,. \label{classIV3}
\ee 
According to the results of the previous section, in the limit $\omega\rightarrow 
\infty$ this spacetime should reduce to a solution of GR with the same symmetries but 
sourced by a canonical, minimally coupled scalar field $\Phi(r)$ which must coincide 
with the 
Einstein frame scalar $\tilde{\phi}(r)$. But the general spherical, static, 
asymptotically flat solution of the Einstein equations sourced by a free scalar 
field is well known: it is the Fisher-Janis-Newman-Winicour-Buchdahl-Wyman 
(FJNWBW) solution hosting a central naked singularity \cite{Fisher,Wyman}. 
Let us check if our statement based on Sec.~\ref{sec:2} is true. By taking 
the limit $\omega\rightarrow \infty$, the Brans-Dicke line 
element~(\ref{new1})  reduces to   
\begin{eqnarray}
ds^2_{(\infty)} &=& -\mbox{e}^{ \alpha_0/r} dt^2 + \mbox{e}^{-\alpha_0/r}
\left( \frac{ \gamma/r}{\sinh( \gamma/r)}\right)^4 dr^2 \nonumber\\
&&\nonumber\\
&\, &  +  \mbox{e}^{-\alpha_0/r} \left( \frac{ \gamma/r}{\sinh( 
\gamma/r)}\right)^2   r^2 d\Omega_{(2)}^2 \,.\label{FJNWBW}
\end{eqnarray}
This is indeed recognized as the FJNWBW geometry \cite{Fisher,Wyman}.

The Jordan frame scalar field $\phi(r)=\phi_0 \, \mbox{e}^{-\beta_0/r}  \rightarrow 
\phi_0=$~const. as $\omega\rightarrow \infty$. However, the scalar field 
sourcing  the FJNWBW geometry in GR is \cite{Fisher,Wyman} 
\be
\Phi (r) = \frac{\Phi_*}{r} \,, \;\;\;\;\;\;\;
\Phi_*=\frac{-\sigma}{4\sqrt{\pi}} \label{F1}
\ee
and is exactly the Einstein frame cousin of $\phi(r)$, {\em i.e.}, $ \Phi (r) 
=\tilde{\phi}(r)$ \cite{ValerioFaycalThomasAdriana}.

Let us consider now the case $\gamma=0$, in which the Brans-Dicke solution 
is the Brans class~IV geometry (\ref{classIV1})-(\ref{classIV3}) 
\cite{Brans}. It is well known that, in the $\omega\rightarrow \infty$ 
limit, this solution does not reduce to the corresponding vacuum solution 
of the Einstein equations with the same symmetries, which is the 
Schwarzschild solution, and it cannot describe a black 
hole but only a wormhole throat or a naked singularity 
\cite{PrainFaraoniZambrano, ourBrans} (indeed, the failure of the Brans 
solutions \cite{Brans} to reproduce the Schwarzschild spacetime was one of 
the first occurrences prompting investigation of the anomaly in the GR 
limit of 
Brans-Dicke theory \cite{BanerjeeSen97}). It is easy to see that the 
$\omega \rightarrow \infty$ limit of 
Eqs.~(\ref{classIV1})-(\ref{classIV3}) 
produces $B\rightarrow \alpha_0/2$, $C\rightarrow 0$, and the Yilmaz 
geometry 
\cite{Yilmaz, expometric}
\be
ds_{(\infty)}^2 = -\, \mbox{e}^{ \alpha_0/r} dt^2 +\mbox{e}^{-\alpha_0/r} \left( 
dr^2 +r^2 d\Omega_{(2)}^2 \right) \,,
\ee
which is indeed the Einstein frame counterpart of Brans class~IV 
\cite{ValerioFaycalThomasAdriana}. The Jordan frame scalar field of Brans 
class~IV $\phi(r)=\phi_0 \, \mbox{e}^{-BC/r} \rightarrow 
\phi_0=$~const. but the scalar field sourcing the Yilmaz geometry is instead
\be
\Phi(r)= \frac{\Phi_0}{r} \,,
\ee 
which coincides with the Einstein frame counterpart of $\phi(r)$, {\em i.e.}, 
$\Phi(r)=\tilde{\phi}(r)$ again.

%%%%%%%%%%%%%
\section{The effective fluid approach}
\label{sec:4}
\setcounter{equation}{0}

Let us discuss now an independent approach to the problem of the anomaly 
in the GR limit of BD theory. In this approach, the Brans-Dicke field 
equations are rewritten as effective Einstein equations with the terms 
dependent on $\phi$ and its derivatives moved to the right hand side to 
provide an extra source (in addition to the real matter stress-energy 
tensor $T_{ab}^{(m)}$). In 
the rest of this article we consider only vacuum or conformally invariant 
matter with $T^{(m)}=0$.  Since we want to define  an effective fluid 
with 4-velocity proportional to the gradient of $\phi$, this 4-vector 
 must necessarily be timelike and this section, unlike the previous 
one, is restricted to situations in which $\nabla^c\phi$ is always 
timelike.

\subsection{Imperfect fluid equivalent of the Brans-Dicke field}

When the gradient $\nabla^a \phi$ is 
timelike, one introduces  the fluid 4-velocity  
\be
u^a  = \frac{\nabla^a  \phi}{\sqrt{ -\nabla^e \phi \nabla_e \phi }}\,, 
\label{4-velocity}
\ee
which is normalized, $u^a u_a=-1$.  This 4-velocity provides a $3+1$ 
splitting of spacetime  into the time direction $u^a$ and the  
3-dimensional space $\Sigma_t$ perceived by the 
comoving observers of the effective fluid, which has the Riemannian metric 
\be
h_{ab} \equiv g_{ab} + u_a u_b \,.
\ee 
${h_a}^b$ is the usual projection operator which satisfies  
\begin{eqnarray}
h_{ab} u^a &=& h_{ab}u^b=0 \,,\\
&&\nonumber\\
{h^a}_b \, {h^b}_c &=& {h^a}_c \,, \;\;\;\;\;\;  {h^a}_a=3 \,.
\end{eqnarray}
The fluid 4-acceleration  
\be
\dot{u}^a \equiv u^b \nabla_b 
u^a 
\ee
is purely spatial, $\dot{u}^c u_c=0$.  The  projection of the velocity 
gradient onto the 3-space $\Sigma_t$ is the purely spatial tensor 
\be
V_{ab} \equiv  {h_a}^c \, {h_b}^d \, \nabla_d u_c  \,, \label{Vab}
\ee 
which is decomposed according to
\be
V_{ab}=  \theta_{ab} +\omega_{ab} =\sigma_{ab} +\frac{\theta}{3} \, 
h_{ab}+ \omega_{ab} \,,
\ee
where $\theta_{ab}=V_{(ab)}$ (expansion tensor) is the symmetric part of 
$V_{ab}$, $\theta\equiv {\theta^c}_c =\nabla^c u_c $ is its trace,
\be 
\sigma_{ab} \equiv \theta_{ab}-\frac{\theta}{3}\, h_{ab}
\ee
(shear tensor) is the trace-free part of $\theta_{ab}$, and the vorticity 
tensor $\omega_{ab}=V_{[ab]}$ vanishes identically because the fluid is 
generated by a scalar and $u^a$ is irrotational.  The tensors $h_{ab}$, 
$V_{ab}$, expansion $\theta_{ab}$, and shear $\sigma_{ab}$ are purely 
spatial,
\be
\theta_{ab}u^a = \theta_{ab}u^b =  
  \sigma_{ab}u^a = \sigma_{ab} u^b = 0 \,,
\ee
and ${\sigma^a}_a=0$.  In general, it is \cite{Ellis71}  
\be
\nabla_b u_a =   
\sigma_{ab}+\frac{\theta}{3} \, h_{ab} +\omega_{ab} -  \dot{u}_a 
u_b  =V_{ab} -\dot{u}_a u_b \,. \label{ecce}
\ee
The projection of this equation onto the time direction produces 
$\dot{u}_a$, while the projection onto the 3-space orthogonal to 
$u^a$ gives $V_{ab}$. In our particular case~(\ref{4-velocity}), we have 
\cite{FaraoniCote2018}
\be
h_{ab}= g_{ab}-\frac{ \nabla_a\phi \nabla_b \phi}{ 
\nabla^e\phi \nabla_e\phi}  \,,
\ee
\be
\nabla_b u_a = \frac{1}{ \sqrt{ -\nabla^e\phi \nabla_e \phi}} \left( 
\nabla_a \nabla_b \phi -\frac{  \nabla_a \phi \nabla^c \phi \nabla_b 
\nabla_c \phi}{\nabla^e\phi \nabla_e \phi} \right) \,,
\ee
\begin{eqnarray}
\dot{u}_a &=& \left( -\nabla^e \phi \nabla_e \phi \right)^{-2} 
\nabla^b \phi 
\Big[ (-\nabla^e \phi  \nabla_e \phi)  \nabla_a \nabla_b 
\phi  \nonumber\\
&&\nonumber\\
&\, &   + \nabla^c  \phi \nabla_b \nabla_c \phi \nabla_a 
\phi \Big] \,. \label{acceleration}
\end{eqnarray}
The kinematic quantities of the effective fluid are given in 
Ref.~\cite{FaraoniCote2018}, while the effective stress-energy tensor of 
$\phi$ is described by 
\begin{eqnarray}
8\pi T_{ab}^{(\phi)} &=& \frac{\omega}{\phi^2} \left( \nabla_a \phi 
\nabla_b \phi - 
 \frac{1}{2} \, g_{ab} \nabla^c \phi \nabla_c \phi  \right) \nonumber\\
&&\nonumber\\
&\, & + 
 \frac{1}{\phi} \left( \nabla_a \nabla_b \phi -g_{ab} \square \phi \right) 
- \frac{V}{2 \phi} \, g_{ab}  \label{BDemt}
\end{eqnarray}
and it is covariantly conserved, together with that of ordinary matter
\be
\nabla^b T_{ab}^{(m)}=0 \,, \;\;\;\;\;
\nabla^b T_{ab}^{(\phi)}=0 \,.
\ee
The  effective stress-energy  
tensor $T_{ab}^{(\phi)}$  can be written \cite{Pimentel89,FaraoniCote2018} 
in the imperfect fluid form 
\be 
T_{ab} = \rho u_a u_b + q_a u_b + q_b u_a + \Pi_{ab} 
\,,\label{imperfectTab}
\ee 
where 
\begin{eqnarray}
\rho &=& T_{ab} u^a u^b \,,  \label{rhophi}\\
&&    \nonumber\\
q_a & =&  -T_{cd} \, u^c {h_a}^d \,, \label{qphi}\\
&&  \nonumber\\
 \Pi_{ab} &\equiv & Ph_{ab} + \pi_{ab} = T_{cd} \, {h_a}^c \, {h_b}^d \,, 
\label{Piphi}\\
&&  \nonumber\\
    P &=& \frac{1}{3}\, g^{ab}\Pi_{ab} =\frac{1}{3} \, h^{ab} T_{ab} \,, 
\label{Pphi}\\
&&  \nonumber\\
    \pi_{ab} &=& \Pi_{ab} - Ph_{ab} \,, \label{piphi}
\end{eqnarray}
are the effective energy density, heat flux density, 
stress tensor,   
isotropic  pressure, and anisotropic stresses (the trace-free part 
$\pi_{ab}$ of 
the stress tensor $\Pi_{ab}$) in the comoving frame. In 
this frame, we have 
\be 
q_c u^c = \Pi_{ab} u^b=\pi_{ab} u^b = \Pi_{ab} u^a=\pi_{ab} u^a = 0 \, 
, \,\,\,\,\,\,\,  {\pi^a}_a = 0 \,. 
\ee 
These effective quantities were computed explicitly in  
Refs.~\cite{Pimentel89, FaraoniCote2018}, obtaining: 
\begin{widetext}
\begin{eqnarray}
8 \pi \rho^{(\phi)} &=&  -\frac{\omega}{2\phi^2} \, \nabla^e \phi \nabla_e 
\phi  +  \frac{V}{2\phi} + \frac{1}{\phi} \left( \square \phi -  
\frac{  \nabla^a \phi \nabla^b \phi \nabla_a 
\nabla_b \phi}{ \nabla^e \phi  \nabla_e \phi  } \right)  
\,,\label{effdensity}\\
&&\nonumber\\
8 \pi q_a^{(\phi)}   &=& \frac{\nabla^c  \phi \nabla^d \phi}{\phi 
  \left(-\nabla^e \phi \nabla_e \phi \right)^{3/2} } \,  
\Big(  \nabla_d \phi \nabla_c \nabla_a \phi 
- \nabla_a \phi \nabla_c \nabla_d \phi \Big)  \label{eq:q}\\
&&\nonumber\\
&=& -\frac{\nabla^c \phi \nabla_a \nabla_c\phi}{\phi \left( -\nabla^e \phi 
\nabla_e\phi \right)^{1/2}} - \frac{
\nabla^c\phi \nabla^d\phi \nabla_c\nabla_d \phi}{\phi \left( -\nabla^e\phi 
\nabla_e\phi \right)^{3/2} } \, \nabla_a \phi \,,\\
&&\nonumber\\
8 \pi \Pi_{ab}^{(\phi)}  &=&  (-\nabla^e  \phi \nabla_e \phi)^{-1} \left[ 
\left( 
- \frac{\omega}{2\phi^2}\,  \nabla^e \phi \nabla_e \phi - \frac{\square 
\phi}{\phi} - \frac{V}{2 \phi} \right) \Big( \nabla_a \phi \nabla_b \phi -
g_{ab}  \nabla^e \phi  \nabla_e \phi  \Big) \right.  
\nonumber\\
&&\nonumber\\
&\, & \left.  - \frac{\nabla^d 
\phi}{\phi} \left( \nabla_d  \phi \nabla_a \nabla_b \phi - \nabla_b \phi 
\nabla_a \nabla_d \phi - \nabla_a  \phi \nabla_d \nabla_b \phi + 
\frac{  \nabla_a \phi \nabla_b \phi \nabla^c \phi  \nabla_c \nabla_d 
\phi}{ \nabla^e \phi \nabla_e \phi } \right) \right] \\
&&\nonumber\\
&=& \left( -\frac{\omega}{2\phi^2} \, \nabla^c \phi \nabla_c \phi 
-\frac{\Box\phi}{\phi} -\frac{V}{2\phi} \right) h_{ab} +\frac{1}{\phi} \, 
{h_a}^c {h_b}^d \nabla_c \nabla_d \phi \,,  \label{eq:effPi2}\\
&&\nonumber\\
8 \pi P^{(\phi)}  & = &  - \frac{\omega}{2\phi^2} \, \nabla^e \phi 
\nabla_e \phi - 
\frac{V}{2\phi} - \frac{1}{3\phi}  \left( 2\square \phi + 
\frac{\nabla^a \phi \nabla^b \phi \nabla_b \nabla_a \phi }{\nabla^e \phi 
\nabla_e  \phi }  \right) \,, \label{effpressure}\\
&&\nonumber\\
8 \pi \pi_{ab}^{(\phi)}   &=& \frac{1}{\phi \nabla^e \phi \nabla_e 
\phi } 
\left[ \frac{1}{3} \left( \nabla_a  \phi \nabla_b \phi - g_{ab} \nabla^c 
\phi \nabla_c \phi \right) \left(  \square \phi  - 
\frac{ \nabla^c \phi  \nabla^d \phi \nabla_d \nabla_c \phi }{ \nabla^e \phi 
\nabla_e \phi }   
\right) \right. \nonumber\\
&&\nonumber\\
&\, & \left. + \nabla^d \phi \left(  \nabla_d \phi \nabla_a \nabla_b 
\phi - 
\nabla_b \phi \nabla_a \nabla_d  \phi - \nabla_a \phi \nabla_d \nabla_b 
\phi +  
\frac{ \nabla_a \phi \nabla_b \phi  \nabla^c \phi \nabla_c 
\nabla_d \phi }{ \nabla^e \phi \nabla_e \phi } \right) \right] \,. 
\label{piab-phi} 
\end{eqnarray} 
\newpage
\end{widetext}

\subsection{A symmetry of (electro)vacuum Brans-Dicke gravity and of its 
equivalent effective fluid}

In (electro)vacuum, the Brans-Dicke action~(\ref{STaction}) is invariant 
under the 1-parameter group of symmetries \cite{myBDlimit} 
\begin{eqnarray}
&& g_{ab}  \rightarrow  \bar{g}_{ab}=\phi^{2\alpha} \, g_{ab} \,, 
\label{newg}\\
&&\nonumber\\ 
&& \phi \rightarrow \bar{\phi} = \phi^{1-2\alpha} \,, \;\;\;\;\;\; 
\alpha\neq 1/2 \,,\label{newphi}
\end{eqnarray}
provided that the Brans-Dicke parameter and the scalar field potential  
are changed to  
\begin{eqnarray}
\bar{\omega}( \omega, \alpha)&=& \frac{ \omega 
+6\alpha(1-\alpha)}{(1-2\alpha)^2} \,, \label{newomega}\\ 
&&\nonumber\\
\bar{V}( \bar{\phi}) &=& \bar{\phi}^{\frac{-4\alpha}{1-2\alpha}} 
V\left( \bar{\phi}^{\frac{1}{1-2\alpha}} \right)  \label{newV}
\end{eqnarray}
(because $\alpha\neq 
1/2$, the conformal transformation~(\ref{newg}) is completely different 
from the transformation leading to the Einstein frame 
metric~(\ref{metrictransformation})). 
To understand the use of $\alpha$, the idea is that 
one first discovers a 1-parameter symmetry group of
(electro-)vacuum Brans-Dicke theory. This group is parametrized exactly
by the parameter $\alpha$ appearing in the exponents of both metric and
scalar field redefinitions (in different ways). This parameter is well
defined independently of any limit of the theory. Then, we 
discover that the limit $\omega\rightarrow \infty$ of the theory
corresponds to the limit $\alpha\rightarrow 1/2$ and we use this fact
advantageously for the particular  problem at hand.  Is is shown in 
Ref.~\cite{myBDlimit} that 
the operations~(\ref{newg}), (\ref{newphi}) constitute a 1-parameter 
group of symmetries which, thus far, has seen two uses in the literature:  
to generate 
new solutions of the field equations from 
known ones \cite{ValerioDilekShawn}, and to study 
the anomaly in the $\omega\rightarrow \infty$ limit of 
Brans-Dicke gravity \cite{myBDlimit, mycosmobook}. It is this second use that we are 
interested in here.

The relation~(\ref{newomega}) between the parameters $\bar{\omega}$ and 
$\alpha$ can be inverted, obtaining
\be
\alpha=\frac{ 2\bar{\omega}+3 \pm \sqrt{ (2\bar{\omega}+3)( 
2\omega+3)} }{2(2\bar{\omega}+3)} \,,
\ee
and $\alpha \rightarrow 1/2$ as $\bar{\omega} \rightarrow \infty$, hence 
one can trade these two limits and think of obtaining larger and larger 
$\bar{\omega}$ by means of consecutive symmetry 
transformations~(\ref{newg}), (\ref{newphi}). In (electro)vacuo, this  
transformation 
connects theories within an equivalence class and a change of the 
$\omega$-parameter $\omega\rightarrow \bar{\omega}$ simply moves a 
Brans-Dicke theory within this equivalence class \cite{myBDlimit, 
mycosmobook}. This is true also for the $\bar{\omega}\rightarrow \infty$ 
limit, 
which can be seen as a transformation~(\ref{newg}), (\ref{newphi}) leading 
to larger and larger $\bar{\omega}$ and cannot break this restricted 
conformal invariance and move the theory outside of the equivalence class. 
GR, which is not conformally invariant, does not belong to this class and 
it cannot be reproduced\footnote{Quantization breaks this symmetry 
\cite{Pal2016}.} by the $\omega \rightarrow \infty $ limit under these 
circumstances---one needs to first break the conformal invariance and exit 
this class to be able to obtain GR as the limit \cite{myBDlimit, 
mycosmobook}.

Naturally, the symmetry~(\ref{newg}), (\ref{newphi}) corresponds to a 
symmetry of the effective fluid.  Under~(\ref{newg}), 
(\ref{newphi}), the fluid 4-velocity is mapped to  \cite{FaraoniCote2018} 
\begin{eqnarray}
u_c & \rightarrow & \bar{u}_c \equiv \frac{ \bar{\nabla}_c  
\bar{\phi} }{ 
\sqrt{ - \bar{g}^{cd} \bar{\nabla}_c \bar{\phi} \bar{\nabla}_d 
\bar{\phi} }}       
=\phi^{\alpha} u_c \,, \label{u-property}\\
&&\nonumber\\
u^c & \rightarrow & \bar{u}^c=\phi^{-\alpha} u^c \,, \label{u2}
\end{eqnarray}
while $
\bar{g}^{ab} \bar{u}_a  \bar{u}_b  
= g^{ab} u_a u_b = -1 $. 
The effective fluid quantities transform according to 
\cite{FaraoniCote2018} 
\begin{eqnarray}
\bar{T}_{ab}^{(\bar{\phi}) } &=& T_{ab}^{(\phi) }
+ \frac{\alpha}{4\pi \phi} \left[ \frac{(1+\alpha)}{\phi} \, \nabla_a \phi 
\nabla_b \phi  \right. \nonumber\\
&&\nonumber\\
&\, & \left. +\frac{(\alpha-2)}{2\phi} \, 
\nabla^e\phi \nabla_e \phi \, 
g_{ab} -\left( \nabla_a \nabla_b \phi -g_{ab} \Box \phi \right) \right] 
\,,\nonumber\\
&&\\
\bar{\rho}^{( \bar{\phi})} &=&  \phi^{-2\alpha} \left[ (1-2\alpha) 
\rho^{(\phi)}  - \frac{\alpha(3\alpha + \omega)}{8 \pi \phi^2} 
\, \nabla^e \phi \nabla_e \phi  \right. \nonumber\\
&&\nonumber\\
&\, & \left. + \frac{\alpha V}{8 \pi 
\phi}   \right] \,,\\
&&\nonumber\\
\bar{q}_a^{( \bar{\phi})} &=& \left( 1-2\alpha \right) \phi^{-\alpha}\,   
q_a^{(\phi)}  \,,\\
&&\nonumber\\
\bar{P}^{( \bar{\phi})} &=& \phi^{-2 \alpha} \left[ \left( 1-2\alpha 
\right) P^{(\phi)} + 
\frac{\alpha \left(\alpha - \omega -2 \right)}{8 \pi \phi^2}  \, 
\nabla^e \phi \nabla_e \phi \right.\nonumber\\
&&\nonumber\\
&\, & \left.  - \frac{\alpha V}{8 \pi 
\phi}  \right] \,,\\
&&\nonumber\\
\bar{\Pi}_{ab}^{( \bar{\phi})} &=& \left( 1-2\alpha 
\right)\Pi_{ab}^{(\phi)}  \nonumber\\
&&\nonumber\\
&\, & + \frac{\alpha}{8 \pi \phi} \left[ \frac{\left( \alpha - 
\omega-2\right)}{\phi} \, \nabla^e \phi \nabla_e \phi  - V \right] h_{ab} 
\,,\nonumber\\
&&\\
\bar{\pi}_{ab}^{( \bar{\phi})} &=& \left( 1-2\alpha \right) 
\pi_{ab}^{(\phi)}  \,.
\end{eqnarray}
In the limit $\alpha \rightarrow 1/2$ (corresponding to $\bar{\omega} 
\rightarrow \infty$) the imperfect fluid quantities, {\em i.e.}, the heat 
flux $\bar{q}_a^{(\bar{\phi})} $ and the anisotropic stresses 
$\bar{\pi}_{ab}^{(\bar{\phi})}$, which are proportional to 
$\left(1-2\alpha\right)$ vanish identically, but there remain 
non-vanishing contributions to the effective energy density and pressure: 
\begin{eqnarray} 
\bar{T}_{ab}^{(\bar{\phi}) }  \rightarrow 
\bar{T}_{ab}^{(\infty)} &=& \frac{1}{8\pi} \left[ 
\frac{(2\omega+3)}{2\phi^2} \left( \nabla_a \phi \nabla_b \phi
-\frac{1}{2} \, g_{ab} \nabla^c \phi\nabla_c \phi \right) 
\right.\nonumber\\
&&\nonumber\\
&\, & \left. -\frac{V(\phi)}{2\phi} \, 
g_{ab} \right] \,,\\
&&\nonumber\\
\bar{\rho}^{( \bar{\phi})}  \rightarrow 
\bar{\rho}_{(\infty)} &=& \frac{1}{\phi} \left[ - \frac{ (2\omega+3)}{32\pi\phi^2} 
\, \nabla^c \phi\nabla_c \phi  +\frac{V}{16\pi\phi} \right] \,,\\
&&\nonumber\\
\bar{P}^{( \bar{\phi})}  \rightarrow  
\bar{P}_{(\infty)} &=& \frac{1}{\phi} \left[ - \frac{ (2\omega+3)}{32\pi\phi^2} 
\, \nabla^c \phi\nabla_c \phi  -\frac{V}{16\pi\phi} \right] \,.
\end{eqnarray}
By introducing $\tilde{g}_{ab} \equiv \phi g_{ab}$, $\tilde{g}^{ab} \equiv 
\phi^{-1} g^{ab}$, $ \Phi= \sqrt{ \frac{|2\omega+3|}{16\pi}} \,\ln \left( 
\frac{\phi}{\phi_0} 
\right)$ ({\em i.e.}, the Einstein frame metric, inverse metric, and 
scalar field), 
these quantities are written as
\begin{eqnarray} 
\bar{T}_{ab}^{(\infty)} &=&  
\mbox{sign}\left( 2\omega+3 \right) \left[
\tilde{\nabla}_a \Phi \tilde{\nabla}_b \Phi -\frac{1}{2} \, \tilde{g}_{ab} 
\left( 
\tilde{g}^{cd} \tilde{\nabla}_c \Phi \tilde{\nabla}_d \Phi \right) 
\right] \nonumber\\
&&\nonumber\\
&\, & - U(\Phi) \tilde{g}_{ab}  \,, \\
&&\nonumber\\
\bar{\rho}_{(\infty)} &=& -\frac{\mbox{sign}\left( 2\omega+3 
\right)}{2} \,   \tilde{g}^{ab} \tilde{\nabla}_a \Phi \tilde{\nabla}_b  
\Phi  + U(\Phi)  \,,\\
&&\nonumber\\
\bar{P}_{(\infty)} &=& -\frac{ \mbox{sign}\left( 2\omega+3 \right) }{2} \, 
\tilde{g}^{ab}  \tilde{\nabla}_a \Phi \tilde{\nabla}_b 
\Phi   - U(\Phi)   \,.
\end{eqnarray}
The stress-energy tensor reduces to that of a minimally coupled scalar 
field which has the perfect fluid form \cite{MCscalar, Pimentel89}. If 
$2\omega+3>0$, it is a canonical scalar, while if $2\omega+3<0$, it is a 
phantom field.  The imperfect fluid terms (the heat flux density 
$\bar{q}^a$ and the anisotropic stresses $\bar{\pi}_{ab}$)  vanish in the 
limit $\alpha \rightarrow 1/2$, while the second order derivatives of 
$\phi$ (which cause the effective stress-energy tensor of the Brans-Dicke 
field to deviate from the canonical form) are all contained in the terms 
$\rho^{(\phi)}$ and $P^{(\phi)}$ which, being weighted by a factor 
$(1-2\alpha)$, vanish in this limit. What is more, if $2\omega+3>0$, the 
canonical and minimally coupled scalar field $\Phi$ left in the limit is 
nothing but the Einstein frame scalar corresponding to the Jordan frame 
$\phi$, {\em i.e.}, $\Phi=\tilde{\phi}$. However, the Einstein frame was 
not used in any way in the considerations of this section. This result 
reproduces that of Sec.~\ref{sec:2} using an independent and complementary 
method, but is limited by the requirement that the gradient $\nabla^c 
\phi$ be timelike.

As a final comment, consider the Brans-Dicke field equation~(\ref{BDfe2}) 
for the scalar field which, in the presence of conformal matter (including 
(electro)vacuum), and for $\omega=$~const. and $V(\phi) \equiv 0$, reduces 
to $\Box \phi=0$. Since this equation does not contain explicitly the 
parameter $\omega$, its form is not affected by the limit $\omega 
\rightarrow \infty$ and, indeed, the scalar field $\Phi$ left behind in 
the $\omega \rightarrow \infty$ limit solves the same formal equation 
$\Box \Phi=0$ (although the metric and its covariant derivative operator 
change during the limit).

\section{Conclusions}
\label{sec:5}
\setcounter{equation}{0}

When considering particular solutions of Brans-Dicke theory, one should 
keep in mind that the limit of the metric tensor taken in a specific 
coordinate system may not be unique \cite{Geroch}. A possible, rigorous 
way to take the limit of a spacetime as a parameter varies consists of the 
geometric, coordinate-invariant method of Refs.~\cite{MacCallum}. The 
two methods employed in the present work are also covariant.

We have studied the limit of (electro)vacuum Brans-Dicke gravity with no 
scalar field potential as the Brans-Dicke coupling parameter $\omega $ 
becomes infinite.  The first method, based on an analysis of Einstein 
frame quantities, elucidates the relation between the two GR relatives of 
Jordan frame spacetimes in Brans-Dicke theory, the Einstein frame 
formulation and the Jordan frame limit $\omega\rightarrow \infty$. The 
second method combines an effective fluid description of Brans-Dicke 
gravity (in which the Jordan frame scalar $\phi$ is equivalent to an 
imperfect fluid sourcing effective Einstein equations)  with a 1-parameter 
symmetry group of the theory \cite{myBDlimit}. The two methods are 
independent of each other and complementary and produce the same result. 
As $\omega\rightarrow \infty$, the metric $g_{ab}$ of an (electro)vacuum 
Brans-Dicke spacetime $\left(M, g_{ab}, \phi \right)$ (where $M$ is the 
spacetime manifold) tends to a metric which does not solve the 
(electro)vacuum Einstein equations. It solves instead the Einstein 
equations sourced by a minimally coupled scalar field. Moreover, this 
scalar coincides with the Einstein frame scalar field $\tilde{\phi}$ 
corresponding to the Jordan frame scalar $\phi$ present before taking any 
limit. Contrary to much reasoning in the literature, the methods used do 
not rely crucially on assuming {\em a priori} 
the asymptotics of the Brans-Dicke scalar field as $\omega\rightarrow 
\infty$. The formal explanation of the failure of the $\omega\rightarrow 
\infty$ limit of (electro)vacuum Brans-Dicke theory to reproduce a GR 
solution with the same matter source, proposed long ago in 
Refs.~\cite{myBDlimit}, is put on a more physical basis by the present 
analysis. Essentially, the $\omega\rightarrow \infty$ limit does not 
freeze the scalar gravitational degree of freedom of Brans-Dicke theory 
but demotes it to the role of an ``ordinary'' minimally coupled scalar 
field with stress-energy tensor quadratic in the first derivatives of this 
field and no contribution linear in the second derivatives. This is still 
a non-trivial dynamical field, but it has changed its gravitational nature 
to that of a matter field.

As an example, we have performed a test of our conclusions using the 
general static, spherical, asymptotically flat solution of the vacuum 
Brans-Dicke equations \cite{Just, Bronnikov, ValerioFaycalThomasAdriana} 
(excluding black holes, for which the Brans-Dicke scalar is trivial 
outside the event horizon \cite{nohair}). The result, which produces the 
FJNWBW metric hosting a central naked singularity, confirms the 
conclusions of Sec.~\ref{sec:2} and Sec.~\ref{sec:4}.

Shedding light on this problem that is two decades old is important for 
three reasons. First, contrary to two decades ago, there 
is now a major 
experimental effort to test, or constrain, gravity at many scales 
involving cosmology \cite{Koyama, Euclid}, supermassive black holes 
\cite{smBHtests}, binary systems of compact objects emitting gravitational 
waves \cite{gwtests}, and the Solar System \cite{BertottiIessTortora, 
Will, tests, Padilla, Psaltis}. Since the PPN formalism based on the 
weak-field limit of gravitational theories \cite{Will} is the basis for 
many of these analyses, one should worry about the exact meaning of this 
formalism if vacuum Brans-Dicke solutions (which describe wormhole throats 
or naked singularities) do not go over to the corresponding GR solutions 
(usually, the Schwarzschild geometry which describes a black hole). 
Second, a clear picture of the anomalous limit of (electro-)vacuum 
Brans-Dicke theory to GR may help understanding dynamical 
attractor mechanisms of scalar-tensor to GR gravity. Finally,
in the context of the thermodynamics of spacetime \cite{Jacobson1, 
Jacobson2}, modified theories of gravity (including scalar-tensor gravity) 
could correspond to deviations from an equilibrium state corresponding to 
Einstein theory \cite{ChircoLiberati}. The implications of the new picture 
of the $\omega\rightarrow \infty$ limit of (electro)vacuum Brans-Dicke 
theory for these two areas of research will be discussed in future 
publications.

\begin{acknowledgments}

V.F. is grateful to Lorne Nelson and Shawn Belknap-Keet for discussions 
and to Thibault Damour for pointing out Ref.~\cite{Just}. This work is 
supported by the Natural Sciences and Engineering Research Council of 
Canada (Grant No.~2016-03803) and by Bishop's University.

\end{acknowledgments}

 \end{document}